\documentclass[conference]{IEEEtran}
\usepackage{balance}
\usepackage[T1]{fontenc}
\usepackage[utf8]{inputenc}
\usepackage{amssymb}
\usepackage[pdftex]{graphicx}
\usepackage{mathtools}
\usepackage{mathabx}
\usepackage{array}
\usepackage{mathtools}
\usepackage[noadjust]{cite}
\usepackage{glossaries}
\usepackage{algorithm}
\usepackage{algpseudocode}
\usepackage{xcolor}
\usepackage{tikz}
\usepackage{pgfplots}
\usepgfplotslibrary{fillbetween} 

\makeatletter
\let\MYcaption\@makecaption
\makeatother

\usepackage[font=footnotesize]{subcaption}

\makeatletter
\let\@makecaption\MYcaption
\makeatother
\algnewcommand{\LineComment}[1]{\(\triangleright\) #1}

\newacronym{ACM}{ACM}{adaptive coding and modulation}
\newacronym{ADC}{ADC}{analog-to-digital conversion}
\newacronym{AGC}{AGC}{automatic gain control}
\newacronym{AWGN}{AWGN}{additive white Gaussian noise}
\newacronym{BER}{BER}{bit error rate}
\newacronym{BLER}{BLER}{block error rate}
\newacronym{BP}{BP}{backpropagation}
\newacronym{BPTT}{BPTT}{backpropagation through time}
\newacronym{CE}{CE}{cross-entropy}
\newacronym{CFO}{CFO}{carrier frequency offset}
\newacronym{CSI}{CSI}{channel state information}
\newacronym{DAC}{DAC}{digital-to-analog conversion}
\newacronym{DL}{DL}{deep learning}
\newacronym{DFT}{DFT}{discrete Fourier transform}
\newacronym{FFT}{FFT}{fast Fourier transform}
\newacronym{GAN}{GAN}{generative adversarial network}
\newacronym{GRU}{GRU}{gated recurrent unit}
\newacronym{iid}{i.i.d.\@}{independent and identically distributed}
\newacronym{IFFT}{IFFT}{inverse fast Fourier transform}
\newacronym{KL}{KL}{Kullback-Leibler}
\newacronym{LSTM}{LSTM}{long short-term memory}
\newacronym{MDP}{MDP}{Markov decision process}
\newacronym{ML}{ML}{machine learning}
\newacronym{MLP}{MLP}{multilayer perceptron}
\newacronym{MIMO}{MIMO}{multiple-input multiple-output}
\newacronym{MSE}{MSE}{mean squared error}
\newacronym{NN}{NN}{neural network}
\newacronym{OFDM}{OFDM}{orthogonal frequency-division multiplexing}
\newacronym{pdf}{pdf}{probability density function}
\newacronym{pmf}{pmf}{probability mass function}
\newacronym{PSNR}{PSNR}{Peak Signal to Noise Ratio}
\newacronym{RBF}{RBF}{Rayleigh block-fading}
\newacronym{ReLU}{ReLU}{rectified linear unit}
\newacronym{RL}{RL}{reinforcement learning}
\newacronym{RNN}{RNN}{recurrent neural network}
\newacronym{SFO}{SFO}{sampling frequency offset}
\newacronym{SNR}{SNR}{signal-to-noise ratio}
\newacronym{SINR}{SINR}{signal-to-interference-plus-noise ratio}
\newacronym{SGD}{SGD}{stochastic gradient descent}
\newacronym{wrt}{w.r.t.\@}{with respect to}

\renewcommand{\vec}[1]{\mathbf{#1}}
\newcommand{\vecs}[1]{\boldsymbol{#1}}

\newcommand{\bv}{\vec{b}}

\newcommand{\lv}{\vec{l}}
\newcommand{\mv}{\vec{m}}

\newcommand{\pv}{\vec{p}}

\newcommand{\rv}{\vec{r}}

\newcommand{\uv}{\vec{u}}

\newcommand{\wv}{\vec{w}}
\newcommand{\xv}{\vec{x}}
\newcommand{\yv}{\vec{y}}

\newcommand{\thetav}{\vecs{\theta}}
\newcommand{\psiv}{\vecs{\psi}}

\newcommand{\Id}{\vec{I}}

\newcommand{\Pm}{\vec{P}}

\newcommand{\Sm}{\vec{S}}

\newcommand{\Wm}{\vec{W}}
\newcommand{\Xm}{\vec{X}}
\newcommand{\Ym}{\vec{Y}}

\newcommand{\Ac}{{\cal A}}

\newcommand{\Cc}{{\cal C}}

\newcommand{\Lc}{{\cal L}}

\newcommand{\Nc}{{\cal N}}

\newcommand{\Sc}{{\cal S}}
\newcommand{\Tc}{{\cal T}}

\newcommand{\CC}{\mathbb{C}}
\newcommand{\MM}{\mathbb{M}}

\newcommand{\RR}{\mathbb{R}}

\newcommand{\tp}{^{\mathsf{T}}}

\newcommand{\LB}{\left(}
\newcommand{\RB}{\right)}
\newcommand{\LP}{\left\{}
\newcommand{\RP}{\right\}}
\newcommand{\LSB}{\left[}
\newcommand{\RSB}{\right]}

\renewcommand{\log}[1]{\mathop{\mathrm{log}}\LB #1\RB}
\renewcommand{\exp}[1]{\mathop{\mathrm{exp}}\LB #1\RB}

\newcommand{\EE}{{\mathbb{E}}}

 \newcommand{\argmin}[1]{\underset{#1}{\operatorname{arg}\,\operatorname{min}}\;}

\begin{document}
\title{End-to-End Learning of Communications Systems Without  a Channel Model}
\author{
\IEEEauthorblockN{Fayçal Ait Aoudia and Jakob Hoydis}
\IEEEauthorblockA{Nokia Bell Labs, Paris-Saclay, 91620 Nozay, France\\\{faycal.ait\_aoudia, jakob.hoydis\}@nokia-bell-labs.com
}}
\maketitle

\begin{abstract}
The idea of end-to-end learning of communications systems through \gls{NN}-based autoencoders has the shortcoming that it requires a differentiable channel model. 
 We present in this paper a novel learning algorithm which alleviates this problem. The algorithm iterates between supervised training of the receiver and \gls{RL}-based training of the transmitter.
 We demonstrate that this approach works as well as fully supervised methods on \gls{AWGN} and \gls{RBF} channels.
 Surprisingly, while our method converges slower on \gls{AWGN} channels than supervised training, it converges faster on \gls{RBF} channels.
 Our results are a first step towards learning of communications systems over any type of channel without prior assumptions. 
\end{abstract}
\glsresetall

\section{Introduction}
End-to-end learning of communications systems is a fascinating novel concept \cite{8054694} whose goal is to learn full transmitter and receiver implementations  which are optimized for a specific performance metric and channel model. This can be achieved by representing transmitter and receiver as \glspl{NN} and by interpreting the whole system as an \emph{autoencoder} \cite{Goodfellow-et-al-2016-Book} which can be trained in a supervised manner using \gls{SGD}. 
Although a theoretically very appealing idea, its biggest drawback hindering practical implementation is that a channel model or, more precisely, the gradient of the instantaneous channel transfer function, must be known. For an actual system, this is hardly the case since the channel is generally a black box for which only inputs and outputs can be observed. A simple work-around proposed in \cite{dorner2017deep} consists in fine-tuning of the receiver based on measured data after initial learning on a channel model. However, with this approach, the transmitter cannot be fine-tuned, resulting in sub-optimal performance.

In this work, we investigate if methods from the field of (deep) \gls{RL} can be used to circumvent the problem of a missing channel gradient. In essence, \gls{RL} provides a theoretical foundation for obtaining an estimate of the gradient of an arbitrary loss function \gls{wrt}  actions taken by an agent \cite{sutton1998reinforcement}. In our case, this agent is the transmitter and the loss is a performance metric provided by the receiver.
The main contribution of this paper is to show that knowledge of the channel model and the instantaneous channel transfer function is indeed not needed. This implies that the autoencoder can be trained from pure observations alone without any knowledge of the underlying channel model. The key to achieve this is the well-known technique of \emph{policy learning} \cite{NIPS1999_1713}. Inspired by this technique, we develop a novel algorithm for end-to-end training which iterates between two phases: (i) supervised training of the receiver and (ii) \gls{RL}-based training of the transmitter based on an estimated gradient of the loss. Comparison with the fully supervised training method of \cite{8054694} on \gls{AWGN} and \gls{RBF} channels reveals essentially identical performance.
Although our method requires considerably more training iterations on \gls{AWGN} channels, it converges faster on \gls{RBF} channels. This is surprising as our scheme relies on noisy gradient estimates.

The idea of autoencoder-based end-to-end learning of communications systems was pioneered in \cite{osheaae2016, 8054694} and the first proof-of-concept using off-the-shelf software-defined radios was described in \cite{dorner2017deep}. Since then, numerous extensions of the original idea towards channel coding \cite{kim2018communication}, \gls{OFDM} \cite{kimOFDM2018, felix2018ofdm}, and \gls{MIMO} \cite{osheamimo2017} have been made, which all demonstrate the versatility of this approach. Another line of work considers the autoencoder idea for joint source-channel coding \cite{farsad2018text}, but without learning of low-level physical layer tasks. The idea of training an \gls{NN}-based transmitter using a policy gradient was explored in \cite{devrieze2018multiagent} for a non-differentiable receiver which treats detection as a clustering problem. 

\textbf{Notations:}
Boldface upper- and lower-case letters denote matrices and column vectors. $\RR$ and $\CC$ denote the sets of real and complex numbers.
The gradient and Jacobian operators \gls{wrt} the set of parameters $\thetav$ are both denoted $\nabla_{\thetav}$; $(\cdot)\tp$ is the transpose operator.
The complex Gaussian distribution with mean $\mv$ and covariane matrix $\Sm$ is denoted by $\Cc\Nc(\mv,\Sm)$. 

 \section{Learning End-to-end Communications Systems} \label{sec:algo}

 \begin{figure}
    \centering
      \includegraphics[width=\linewidth]{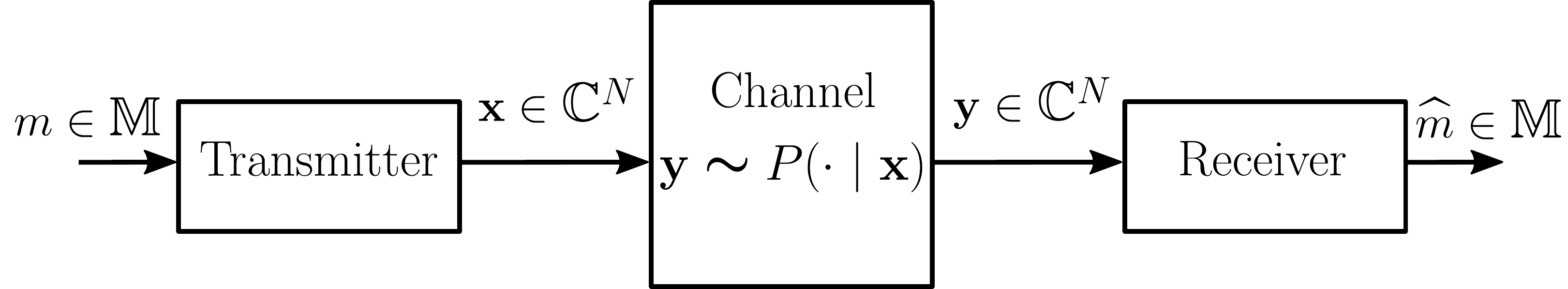}
    \caption{A typical communications system operating over an unknown channel. End-to-end learning describes the process of jointly optimizing the transmitter and the receiver \gls{wrt} a specific performance metric.
}
\label{fig:big_pic}
\vspace*{-10pt}
\end{figure}

 A point-to-point communications system consists of two nodes that aim to reliably exchange information over a channel. The channel acts as a stochastic system, whose output $\yv$ follows a probability distribution conditional on its input $\xv$, i.e., $\yv \sim P(\yv|\xv)$. The task of the transmitter is to communicate messages $m$ drawn from a finite discrete set $\MM = \{1,\dots,M\}$, while the receiver aims to detect the sent messages from the received signals, as illustrated in Fig.~\ref{fig:big_pic}.

Typically, the design of communications systems relies on dividing the transmitter and the receiver into individual blocks, each performing one task such as modulation or channel coding. However, it is unclear if this component-wise approach enables the best possible performance, and the attempts to jointly optimize these revealed intractable or too computationally complex to be practical~\cite{504941}. This motivates the use of \gls{ML} to enable optimization of communications systems for end-to-end performance, without the need for compartmentalization of transmitter and receiver.

The key idea of the approach proposed in this work is to implement transmitter and receiver as two separate parametric functions that are jointly optimized to meet application specific performance requirements. The transmitter is represented by $f^{(T)}_{\thetav_T}: \MM \to \CC^{N}$, where $N$ is the number of channel uses, and $\thetav_T$ is the set of parameters. The receiver is implemented as $f^{(R)}_{\thetav_R} : \CC^{N} \to \LP \pv \in \RR_+^M~|~\sum_{i=1}^M p_i = 1 \RP$, where $\thetav_R$ is the set of parameters and $\pv$ a probability vector over the messages. The purpose of the receiver is to predict $m$ given $\yv$ by estimating the conditional probability $P(m~|~\yv)$. This is done by learning the conditional log-likelihood estimator~\cite{Goodfellow-et-al-2016-Book} 
\begin{equation}
  \thetav_R^* = \argmin{\thetav_R} L(\thetav_R)
\end{equation}
where $L$ is the \gls{CE} defined as
\begin{equation}\label{eq:ce-loss}
  L(\thetav_R) = \frac{1}{S} \sum_{i=1}^S \underbrace{-\log{\LSB f_{\thetav_R}^{(R)}(\yv^{(i)})\RSB_{m^{(i)}}}}_{l^{(i)}} 
\end{equation}
which assumes that the training examples are \gls{iid}, $S$ is the size of the training set, $m^{(i)}$ is the $i$th training example, $l^{(i)}$ is the \emph{per-example loss}, and $\yv^{(i)}$ is the corresponding received signal.

\subsection{Training Process Overview}
\label{sec:training}

Unlike previous approaches that represent  transmitter, channel, and receiver as a single parametric function during the training~\cite{8054694}, we implement transmitter and receiver  by two different parametric functions that can be independently optimized. While the former approach requires a differentiable channel model that must match the actual channel over which the system is supposed to communicate, the training process we propose here does not require any channel model at all.

Next, we walk through the full training process which is provided in Algorithm~\ref{alg:training}. It is assumed that transmitter and receiver have access to a sequence of training examples $m_{\Tc}^{(i)},~i=1,2,\dots$. This can be achieved through  pseudorandom number generators initialized with the same seed. An iteration of the training algorithm is made of two phases, one for the receiver (Sec.~\ref{subsec:rxtrain}), and one for the transmitter (Sec.~\ref{subsec:txtrain}). This training process is carried out until a stop criterion is satisfied (e.g., a fixed number of iterations, a fixed number of iterations during which the loss has not decreased, etc.). As the training algorithm alternates between training of receiver and transmitter, respectively, it is referred to as the \emph{alternating training algorithm}. The intuition behind this process is that, at each iteration, the receiver is improved for fixed transmitter parameters $\thetav_T$, then the transmitter is improved for  fixed receiver parameters  $\thetav_R$. By iteratively carrying out this process, the end-to-end system should improve. Formal study of the convergence of this approach is under investigation.

We assume that $f^{(T)}_{\thetav_T}$ and $f^{(R)}_{\thetav_R}$ are differentiable \gls{wrt} their parameters which are adjusted through gradient descent on the loss function. The most widely used algorithm to perform this task is \gls{SGD}~\cite{Goodfellow-et-al-2016-Book}, or one of its numerous variants, which iteratively updates the parameters as follows:
\begin{equation}
  \thetav^{(j+1)} = \thetav^{(j)} - \eta \nabla_{\thetav}\widetilde{L}(\thetav^{(j)})
\end{equation}
where $\eta > 0$ is the \emph{learning rate}, and $\nabla_{\thetav} \widetilde{L}$ is an approximation of the loss function gradient.
With \gls{SGD}, the training dataset is sampled at each iteration to constitute a \emph{minibatch}, and the gradient of $L$ is approximated using this minibatch.
\gls{SGD} is used in both phases of the alternating training scheme to optimize the parameters of the transmitter and the receiver. The rest of this section details these two stages.

\begin{algorithm}
\caption{Alternating training algorithm}
\label{alg:training}
\begin{algorithmic}[1]
\State \LineComment{Main loop}
\While {\textrm{Stop criterion not met}}
  \State \Call{TrainReceiver}{\null}
  \State \Call{TrainTransmitter}{\null}
\EndWhile

\State \LineComment{Receiver training}
\Function{TrainReceiver}{\null}
  \Repeat
    \State \LineComment{Transmitter:}
    \State $\mv_{\Tc} \gets \Call{TrainingSource}{B_R}$ \label{lst:td_data_send_s}
    \State $\Xm \gets f^{(T)}_{\thetav_T}(\mv_{\Tc})$
    \State $\Call{Send}{\Xm}$ \label{lst:td_data_send_e}
    \State \LineComment{Receiver:}
    \State $\Ym \gets \Call{Receive}{\null}$ \label{lst:td_data_recv_s}
    \State $\Pm \gets f^{(R)}_{\thetav_R}(\Ym)$ \label{lst:td_data_recv_e}
    \State $\mv_{\Tc} \gets \Call{TrainingSource}{\null}$ \label{lst:td_sgd_s}
    \State $\Call{SGD}{\thetav_R, \mv_{\Tc}, \Pm}$ \label{lst:td_sgd_e}
  \Until{Stop criterion is met}
\EndFunction

\State \LineComment{Transmitter training}
\Function{TrainTransmitter}{\null}
  \Repeat
    \State \LineComment{Transmitter:}
    \State $\mv_{\Tc} \gets \Call{TrainingSource}{B_T}$ \label{lst:te_data_enc_s}
    \State $\Xm \gets f^{(T)}_{\thetav_T}(\mv_{\Tc})$
    \State $\Xm_p \gets \Call{SamplePolicy}{\Xm}$ \label{lst:te_data_enc_e}
    \State $\Call{Send}{\Xm_p}$ \label{lst:te_data_recv_s}
    \State \LineComment{Receiver:}
    \State $\Ym \gets \Call{Receive}{\null}$
    \State $\Pm \gets f^{(R)}_{\thetav_R}(\Ym)$ \label{lst:te_data_recv_e}
    \State $\mv_{\Tc} \gets \Call{TrainingSource}{\null}$ \label{lst:te_lg_s}
    \State $\lv \gets \Call{PerExampleLosses}{\mv_{\Tc}, \Pm}$ \label{lst:te_lg_e}
    \State $\Call{SendPerExampleLosses}{l}$ \label{lst:te_loss_send_s}
    \State \LineComment{Transmitter:}
    \State $\lv \gets \Call{ReceivePerExampleLosses}{\null}$ \label{lst:te_loss_send_e}
    \State $\Call{SGD}{\thetav_T, \psiv, \lv}$ \label{lst:te_opt}
  \Until{Stop criterion is met}
\EndFunction
\end{algorithmic}
\end{algorithm}

\begin{figure*}
  \centering
  \begin{subfigure}{0.38\linewidth}
    \centering
    \includegraphics[width=\linewidth]{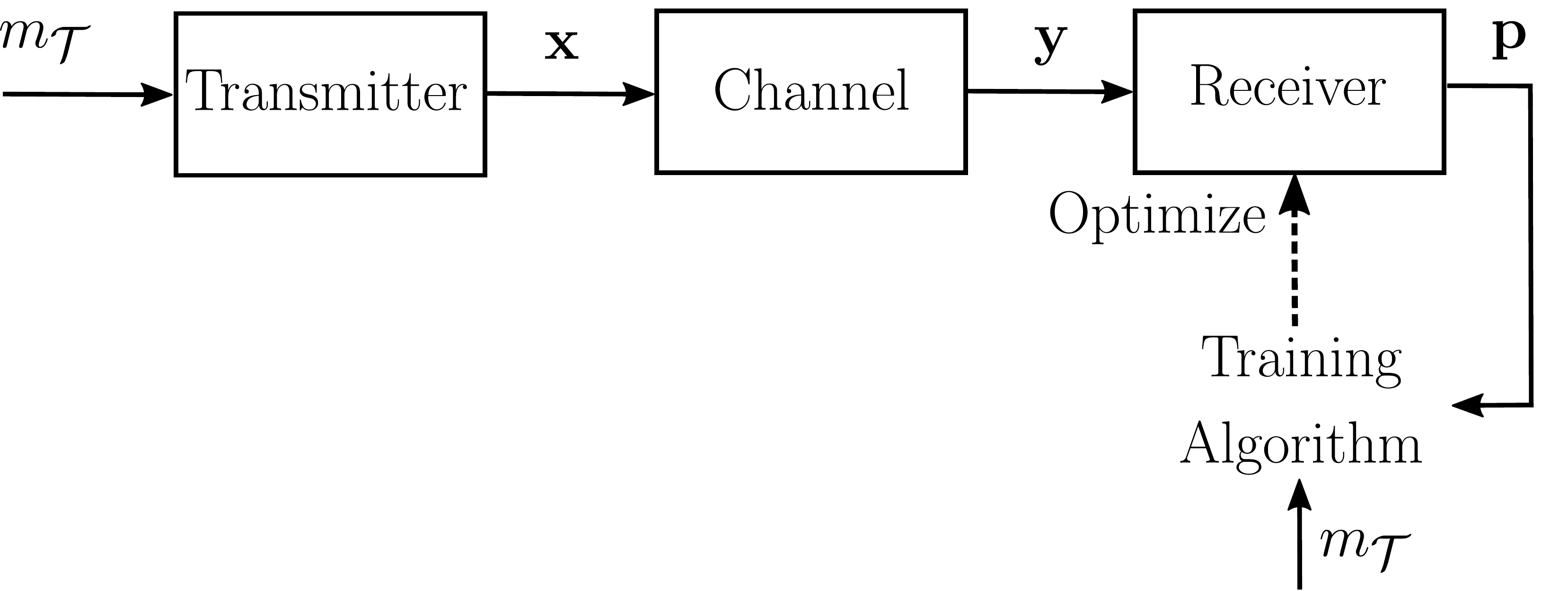}
    \subcaption{Receiver training is a supervised learning task, as the desired outputs are available.}
    \label{fig:train_decoder} 
  \end{subfigure}\hspace{18mm}
  \begin{subfigure}{0.45\linewidth}
    \centering
    \includegraphics[width=\linewidth]{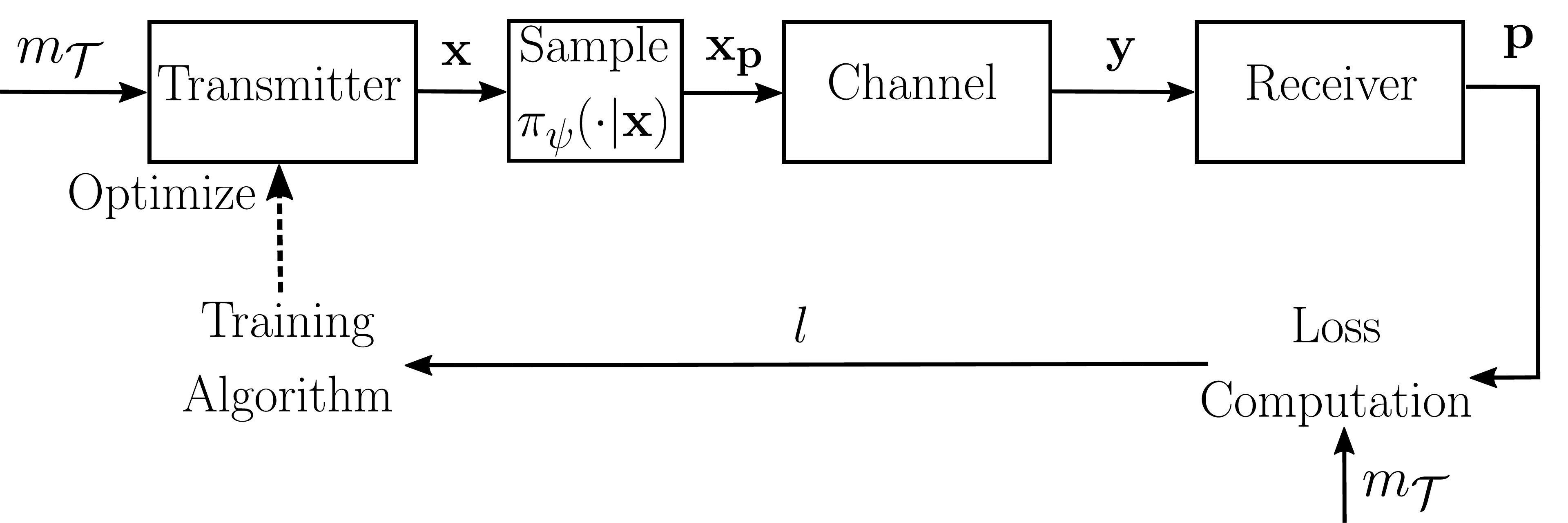}
    \subcaption{Transmitter training is an \gls{RL} task, as the it aims to minimize a scalar loss provided by the receiver.}
    \label{fig:train_encoder}
  \end{subfigure}
  \caption{The two phases of an iteration of the training algorithm: receiver training and transmitter training}
\end{figure*}

\subsection{Receiver Training}\label{subsec:rxtrain}

Receiver training is a supervised learning task, since we assume that the receiver has access to the messages sent for training. The receiver training process is illustrated in Fig.~\ref{fig:train_decoder}, and the pseudocode is given by the function $\textsc{TrainReceiver}$ in Algorithm~\ref{alg:training}.
First, the transmitter generates a minibatch of size $B_R$ of training messages, encodes each training message into $N$ channel symbols, and transmits the minibatch over the channel (lines~\ref{lst:td_data_send_s}--\ref{lst:td_data_send_e}). $\Xm$ is a $B_R$-by-$N$ complex matrix which contains for each example of the minibatch the corresponding complex symbol representation. The receiver obtains the altered symbols $\Ym$, and generates for each training example a probability distribution over $\MM$ (lines~\ref{lst:td_data_recv_s}--\ref{lst:td_data_recv_e}). Finally, an optimization step is performed using \gls{SGD} (or a variant) on the \gls{CE} loss function \eqref{eq:ce-loss} (lines~\ref{lst:td_sgd_s}--\ref{lst:td_sgd_e}).

\subsection{Transmitter Training}\label{subsec:txtrain}

The transmitter's objective is to generate channel symbols that minimize a scalar loss provided by the receiver.
This training scheme corresponds to an \gls{RL} approach. A short background on \gls{RL} is provided in the Appendix. 
The message set $\MM$ corresponds to the state space, and $\CC^{N}$ corresponds to the action space.
To enable exploration, the input of the channel is relaxed to a random variable $\xv_p \sim \pi_{\psiv}(\cdot|\xv)$, which constitutes the stochastic \gls{RL} policy (see Sec.~\ref{sec:eval} for an example). 
The parameter vector $\psiv$ contains parameters specific to the policy distribution, which are only relevant at training time.
When not training the transmitter, $\xv_p = \xv$.
The loss function $\mathcal{L}$ (see Appendix) is an indication of the end-to-end performance and depends on the channel dynamics. $\mathcal{L}$ is only known through the received per-example losses $l$ \eqref{eq:ce-loss} provided by the receiver over an additional reliable channel which is needed during the training process.

The pseudocode of the transmitter training process is shown in the function $\textsc{TrainTransmitter}$ of Algorithm~\ref{alg:training}, and illustrated in Fig.~\ref{fig:train_encoder}.
First, training examples forming a minibatch $\mv_{\Tc}$ of size $B_T$ are encoded into channel symbols $\Xm$.
The stochastic policy is then sampled to generate the channel symbols $\Xm_p$
(lines~\ref{lst:te_data_enc_s}--\ref{lst:te_data_enc_e}).
The channel symbols are sent over the channel, the receiver obtains the altered symbols $\Ym$, and generates for each training example a probability vector over $\MM$ (lines~\ref{lst:te_data_recv_s}--\ref{lst:te_data_recv_e}). Per-example losses $\lv\in\RR^{B_T}$ are then computed based on these vectors and the sent messages $\mv_{\Tc}$ (lines~\ref{lst:te_lg_s}--\ref{lst:te_lg_e}). Next, the per-example losses are sent to the transmitter over a reliable channel, present only during the training (lines~\ref{lst:te_loss_send_s}--\ref{lst:te_loss_send_e}). Finally, an optimization step is performed, using \gls{SGD} or a variant, where the loss gradient is estimated by
\begin{align}
\label{eq:grad_est}
&\nabla_{\thetav_T,\psiv} \widetilde{J}(\mv_{\Tc}, \lv, \Xm_p)\notag\\
&\quad= \frac{1}{B_T} \sum_{i=1}^{B_T} l^{(i)} \nabla_{\thetav_T,\psiv} \log{\pi_{\psiv} \LB \xv_p^{(i)}|f^{(T)}_{\thetav_T}(m_{\Tc}^{(i)}) \RB }.
\end{align}

The number of iterations carried out for training of the transmitter and the receiver at each main iteration of the alternating scheme can either be fixed (as parameters of the algorithm), or can depend on some stop criterion, e.g., stop when no more significant progress is observed.

\section{Generic Transmitter and Receiver Architectures}
Although the alternating training algorithm works for any pair of differentiable parametric functions $ f^{(T)}_{\thetav_T}, f^{(R)}_{\thetav_R}$, we choose to implement them here as \glspl{NN}.
Only feedforward \glspl{NN}, i.e., \glspl{NN} in which connections between \emph{units}, also called neurons, do not form cycles, are considered in this work. A feedforward \gls{NN} of $K$ \emph{layers} is a parametric function $f_{\thetav} : \RR^{N_0} \to \RR^{N_K}$, which maps an input vector $\rv_0 \in \RR^{N_0}$ to an output vector $\rv_K \in \RR^{N_K}$ through $K$ successive layers. Each layer computes an intermediate outcome or \emph{activation vector}
\begin{equation}
  \rv_k = f_{\thetav_k, k}(\rv_{k-1}),\quad k = 1,\dots, K
\end{equation}
where $f_{\thetav_k, k} \to \RR^{N_k}$ is the computation realized by the $k$th layer, and $\thetav_k$ is the set of parameters for this layer. The set of parameters of the entire \gls{NN} is simply the union of all the layers' parameters: $\thetav = \{\thetav_1,\dots,\thetav_K\}$.

\begin{figure}
    \centering
  \begin{subfigure}{0.45\linewidth}
    \centering
    \includegraphics[width=0.76\linewidth]{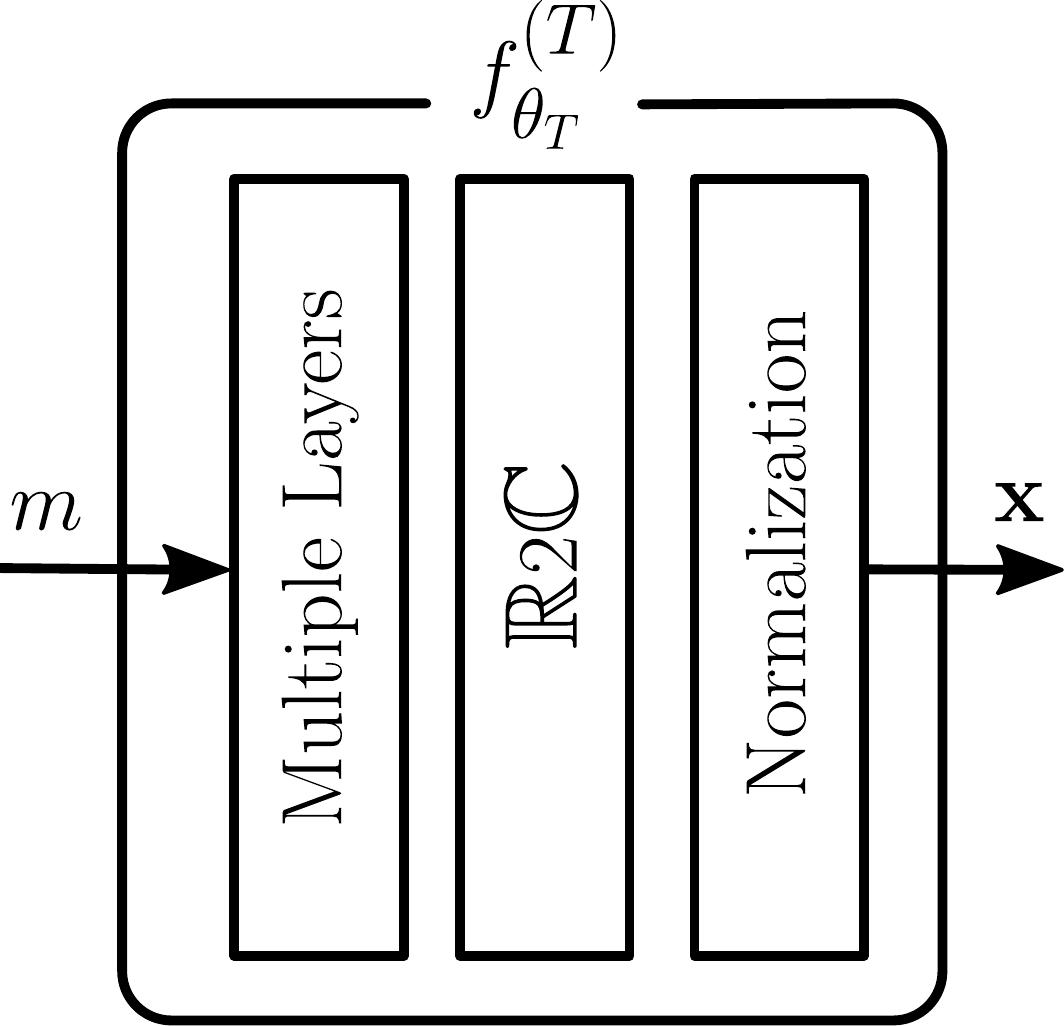}
    \subcaption{Generic transmitter architecture }
    \label{fig:encoder_arch}
  \end{subfigure}\hspace{5mm}
  \begin{subfigure}{0.45\linewidth}
    \centering
    \includegraphics[width=\linewidth]{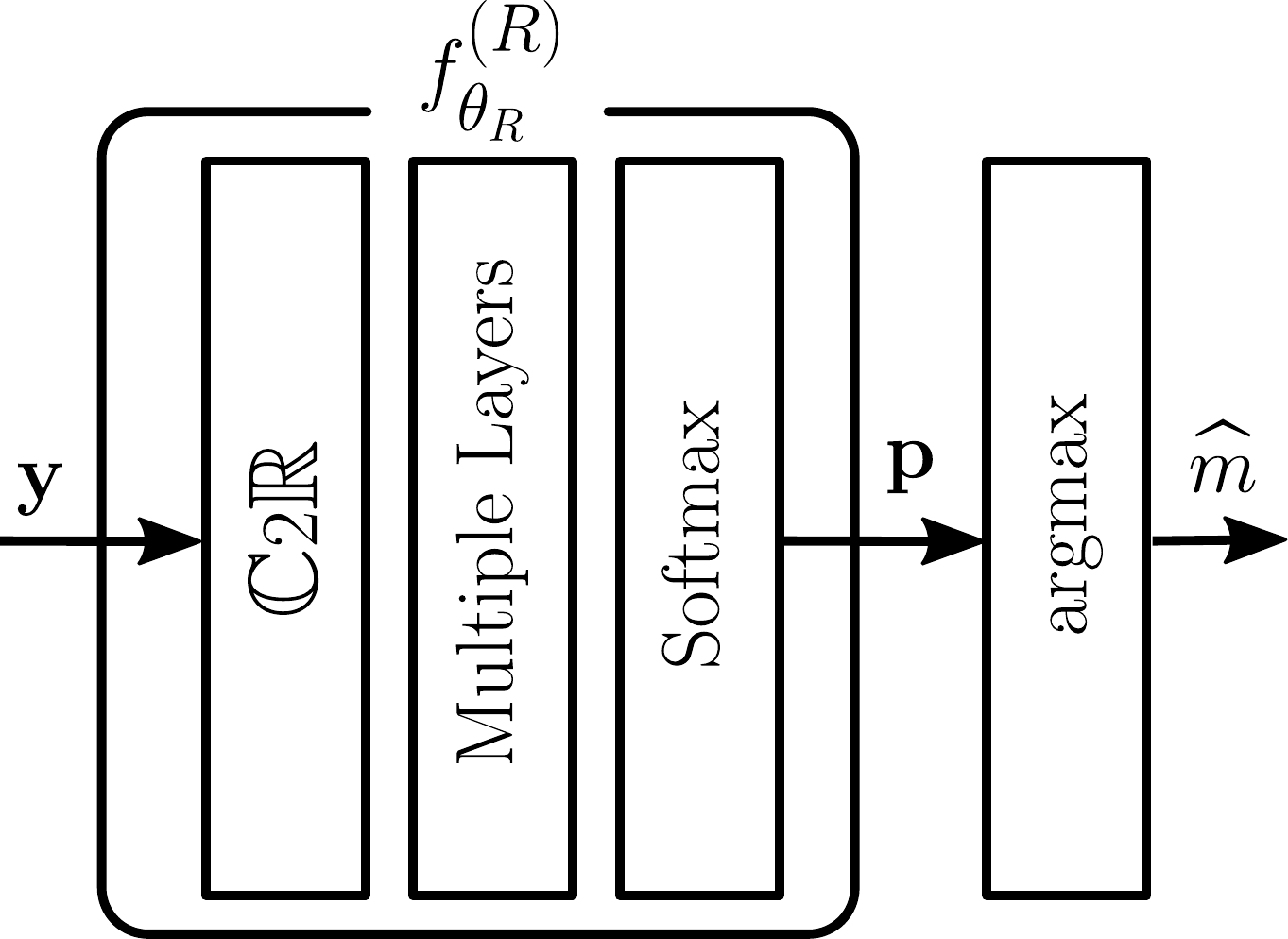}
    \subcaption{Generic receiver architecture}
    \label{fig:decoder_arch}
  \end{subfigure}
  \caption{Generic architectures of the transmitter and receiver}
\end{figure}

The architectures of the transmitter and the receiver can take multiple forms. However, in the context of communications systems, complex baseband symbols are transmitted over the communication channel, and the transmitter must ensure the fulfillment of power constraints.
Therefore, the penultimate layer of the transmitter converts the real outputs of the previous layer into an $N$-dimensional complex-valued vector, and the last layer performs normalization. Normalization guarantees that the average energy per symbol or average energy per message is $1$. The transmitter architecture is shown in Fig.~\ref{fig:encoder_arch}.

The receiver task is to reconstruct the message sent by the transmitter from the received signal. The function $f_{\thetav_R}^{(R)}$ performs soft detection by outputting a probability distribution $\pv$ over $\MM$, and then hard decoding is done by choosing the message with highest probability. The first layer of the receiver converts the received $N$ complex-valued vector $\yv$ into $2N$ real scalars, which are fed to a succession of layers that can be arbitrarily chosen. The last layer of $f_{\thetav_R}^{(R)}$ is a softmax layer to ensure that the output activations form a probability distribution over $\MM$~\cite{Goodfellow-et-al-2016-Book}. Finally, the message with highest probability is chosen as a reconstruction of the sent message.

\section{Evaluation}
\label{sec:eval}
Next, we compare the performance of the proposed alternating training scheme with that of the fully supervised approach of \cite{8054694}, which has been shown to achieve performance close to the best baselines in some scenarios. However, this latter approach relies on the consideration of transmitter, channel, and receiver as a single deep \gls{NN} and, therefore, requires a differentiable model of the channel. This model should at least approximately match the channel on which the communications system is planned to be used. The proposed alternating algorithm does not have such a requirement. Thus, the system can be trained directly over the targeted medium. Evaluations are performed on \gls{AWGN} and \gls{RBF} channels.

\par The \gls{SNR} is defined as
\begin{equation}
  \text{SNR} = \frac{ \EE \LSB \frac1N \lVert\xv \rVert^2_2 \RSB }{\sigma^2}
\end{equation}
where $\EE \LSB \frac1N \lVert\xv \rVert^2_2 \RSB$ is the expected energy per complex symbol, and $\sigma^2$ is the noise variance. The transmitter was set to perform normalization such that $\EE \LSB \frac1N \lVert\xv \rVert^2_2 \RSB$=1, leading to $\text{SNR} = \frac{1}{\sigma^2}.$

\gls{RL} exploration is performed by adding a zero-mean complex normal perturbation $\wv$ to the output of the transmitter, i.e., $\xv_p = \sqrt{1-\sigma_{\pi}^2}\xv + \wv$ where $\wv \thicksim \Cc\Nc(\vec{0}, \sigma_{\pi}^2\Id)$, $\Id$ is the identity matrix of size $N$, and $\sigma_{\pi} \in (0,1)$ is fixed, i.e., not learned during training.
Scaling of the transmitter \gls{NN} output is performed so that the average energy per symbol remains equal to one.
The transmitter \gls{RL} policy $\pi_{\psiv}$ is therefore a \emph{Gaussian policy} with mean $\sqrt{1 - \sigma_{\pi}^2}f^{(T)}_{\thetav_T}$ and covariance matrix $\sigma_{\pi}^2\Id$, so that $\psiv$ is the empty set:
\begin{align}
  &\pi(\xv_p~|~f^{(T)}_{\thetav_T}(m)) =\\
  &\frac{1}{\LB\pi \sigma_{\pi}^2\RB ^N} \exp{-\frac{\lVert\xv_p - \sqrt{1-\sigma_{\pi}^2}f^{(T)}_{\thetav_T}(m)\rVert_2^2}{\sigma_{\pi}^2}}.
\end{align}
This leads to
\begin{align}
  \label{eq:gauss_nabla_log_pol}
  &\nabla_{\thetav_T} \log{\pi(\xv_p~|~m)}\notag\\
   &\quad= \frac{2\sqrt{1-\sigma_{\pi}^2}}{\sigma_{\pi}^2} \LB\nabla_{\thetav_T} f^{(T)}_{\thetav_T}(m)\RB\tp\LB \xv_p - \sqrt{1-\sigma_{\pi}^2}f^{(T)}_{\thetav_T}(m) \RB
\end{align}
which is required to estimate the gradient of the objective function \eqref{eq:grad_est}.
The functions $f^{(T)}_{\thetav_T}$ and $f^{(R)}_{\thetav_R}$ are implemented as deep \glspl{NN} as described in the last section.

Training of the communications systems was done using  the Adam~\cite{Kingma15} optimizer, with an \gls{SNR} set to $10\:$dB ($20\:$dB) for the \gls{AWGN} (\gls{RBF}) channel. The size of $\MM$ was set to $M = 256$, and the number of channel uses $N$ was set to $4$. For the alternating training approach, we used $\sigma_{\pi}^2 = 0.02$.

\subsection{Transmitter and Receiver Architectures}

We implement transmitter and receiver as feedforward \glspl{NN} that leverage only \emph{dense} layers. Dense layers form a simple and widely used type of layer, sometimes also called fully-connected layer. The $k$th layer is defined as
\begin{equation}
  \rv_k = g(\Wm_k \rv_{k-1} + \bv_k)
\end{equation}
where $\Wm_k \in \RR^{N_{k-1}} \times \RR^{N_k}$ is the \emph{weight} matrix, $\bv_k \in \RR^{N_k}$ is the \emph{bias} vector, and $g : \RR \to \RR$ is the \emph{activation function} which is applied elementwise. The trainable parameters of a dense layer are $\thetav_k = \{\Wm_k, \bv_k\}$. The activation function is usually chosen to be nonlinear, which is fundamental to obtain \glspl{NN} that can approximate a wide range of functions.

The transmitter consists of an $M\times M$ embedding with ELU activation functions~\cite{Goodfellow-et-al-2016-Book}, followed by a dense layer of $2N$ units with linear activations. This layer outputs $2N$ reals which are then converted into $N$ complex symbols, and finally normalized as shown in Fig.~\ref{fig:encoder_arch}.

Regarding the receiver, the first layer is a $\CC2\RR$ layer which converts the received $N$ complex symbols into $2N$ real symbols, while the last layer is a dense layer of $M$ units with softmax activations which outputs a probability distribution over $\MM$, as shown in Fig.~\ref{fig:decoder_arch}. In the \gls{AWGN} channel case, a single dense layer with $M$ units and ReLu activation function~\cite{Goodfellow-et-al-2016-Book} was used as hidden layer.

\begin{figure}
    \centering
    \includegraphics[width=0.9\linewidth]{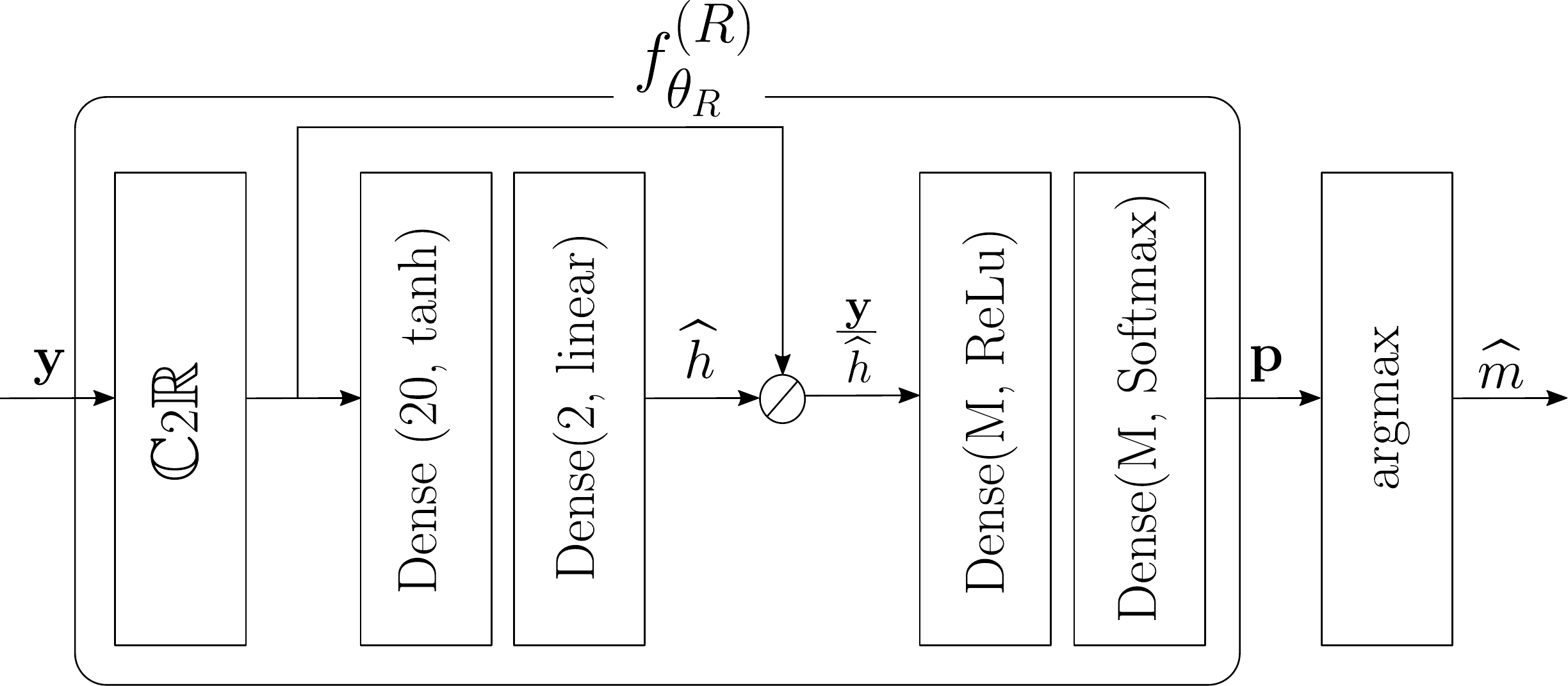}
    \caption{Receiver architecture used for \gls{RBF} channels. After complex-to-real conversion of the received signal, the first two layers estimate the channel response, while the last two layers generate a probability distribution over $\MM$.}
    \label{fig:decoder_rbf}
\end{figure}

In the case of \gls{RBF} channel, using a single dense layer as hidden layer led to poor performance and, therefore, an architecture which incorporates some knowledge about the channel behavior was used \cite[Sec.~III C.]{8054694}. It is well known in \gls{ML} that incorporating expert knowledge through the \gls{NN} architecture can heavily improve the performance of learning systems. Accordingly, the two first hidden layers aim to calculate a value which can be interpreted as an estimate $\widehat{h} \in \CC$ of the channel response $h \in \CC$. The received signal is then divided by this value, and the so obtained signal is fed to a network identical to the one used for the \gls{AWGN} channel, as shown in Fig.~\ref{fig:decoder_rbf}

\subsection{Evaluation Results}

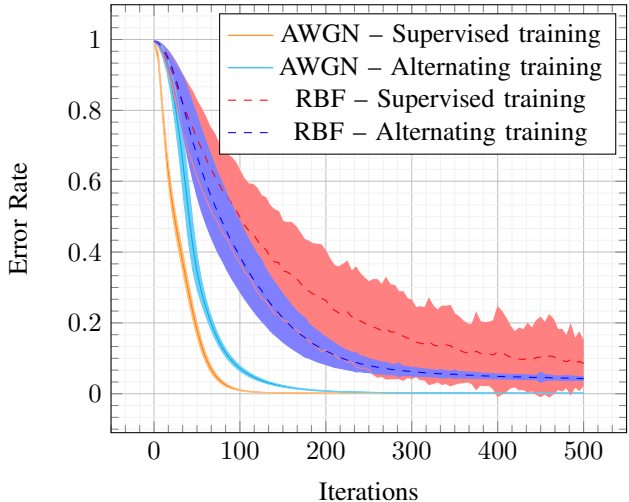
\begin{figure}
    \centering
\begin{tikzpicture}
  \begin{axis}[
    grid=both,
    grid style={line width=.1pt, draw=gray!10},
    major grid style={line width=.2pt,draw=gray!50},
    minor tick num=5,
    xlabel={Iterations},
    ylabel={Error Rate},
]
    \addplot[orange] table [x=iterations, y=awgn_sl, col sep=comma] {figs/training.csv}; \addlegendentry{AWGN -- Supervised training}
    \addplot[orange!50, name path=awgn_sl_up,forget plot] table [x=iterations, y expr=(\thisrowno{1} + \thisrowno{3}), col sep=comma] {figs/training.csv};
    \addplot[orange!50, name path=awgn_sl_down,forget plot] table [x=iterations, y expr=(\thisrowno{1} - \thisrowno{3}), col sep=comma] {figs/training.csv};
    \addplot[orange!50,forget plot] fill between[of=awgn_sl_up and awgn_sl_down];

    \addplot[cyan] table [x=iterations, y=awgn_rl, col sep=comma] {figs/training.csv}; \addlegendentry{AWGN -- Alternating training}
    \addplot[cyan!50, name path=awgn_rl_up,forget plot] table [x=iterations, y expr=(\thisrowno{2} + \thisrowno{4}), col sep=comma] {figs/training.csv};
    \addplot[cyan!50, name path=awgn_rl_down,forget plot] table [x=iterations, y expr=(\thisrowno{2} - \thisrowno{4}), col sep=comma] {figs/training.csv};
    \addplot[cyan!50,forget plot] fill between[of=awgn_rl_up and awgn_rl_down];
    
    \addplot[red, dashed] table [x=iterations, y=rbf_sl, col sep=comma] {figs/training.csv}; \addlegendentry{RBF -- Supervised training}
    \addplot[red!50, name path=rbf_sl_up,forget plot] table [x=iterations, y expr=(\thisrowno{5} + \thisrowno{7}), col sep=comma] {figs/training.csv};
    \addplot[red!50, name path=rbf_sl_down,forget plot] table [x=iterations, y expr=(\thisrowno{5} - \thisrowno{7}), col sep=comma] {figs/training.csv};
    \addplot[red!50,forget plot] fill between[of=rbf_sl_up and rbf_sl_down];
    
    \addplot[blue, dashed] table [x=iterations, y=rbf_rl, col sep=comma] {figs/training.csv}; \addlegendentry{RBF -- Alternating training}
    \addplot[blue!50, name path=rbf_rl_up,forget plot] table [x=iterations, y expr=(\thisrowno{6} + \thisrowno{8}), col sep=comma] {figs/training.csv};
    \addplot[blue!50, name path=rbf_rl_down,forget plot] table [x=iterations, y expr=(\thisrowno{6} - \thisrowno{8}), col sep=comma] {figs/training.csv};
    \addplot[blue!50,forget plot] fill between[of=rbf_rl_up and rbf_rl_down];

\end{axis}

\end{tikzpicture}
    \caption{Evolution of the error rate during the 500 first training iterations}
    \label{fig:training}
\end{figure}

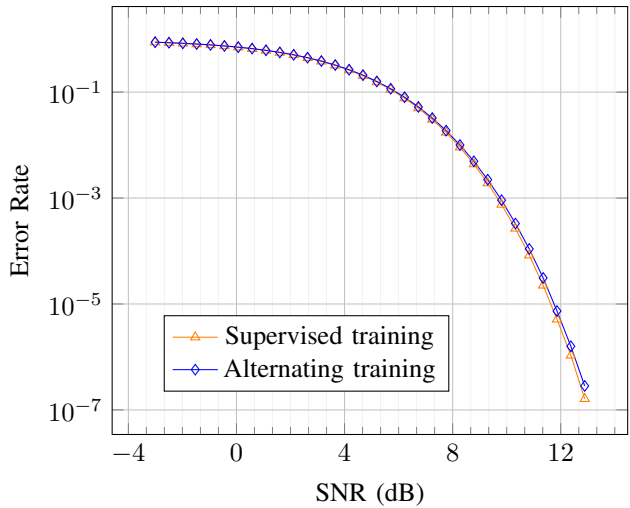
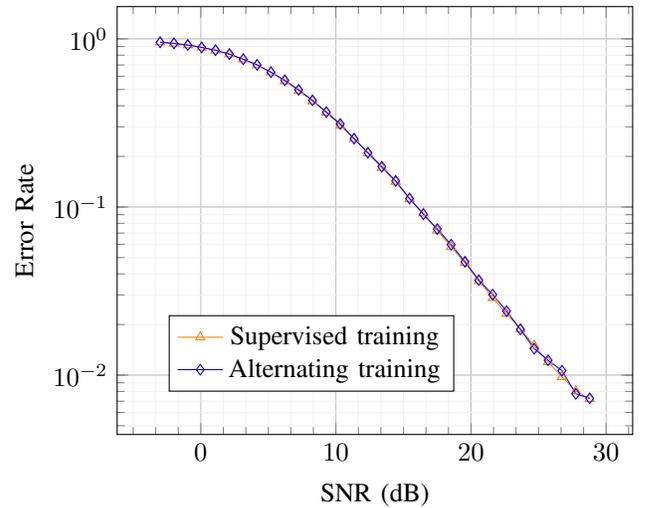
\begin{figure*}
    \centering
    \begin{subfigure}{0.45\linewidth}
	\begin{tikzpicture}
	  \begin{axis}[
	    ymode=log,
	    grid=both,
	    grid style={line width=.1pt, draw=gray!10},
	    major grid style={line width=.2pt,draw=gray!50},
	    minor tick num=5,
	    xlabel={SNR (dB)},
	    ylabel={Error Rate},
	    legend style={at={(0.1, 0.1)},anchor=south west},
	    xtick={-4, 0, 4, 8, 12, 16}
	  ]
	    \addplot[orange, mark=triangle] table [x=snr, y=sl, col sep=comma] {figs/awgn_err_vs_snr.csv};
	    \addplot[blue, mark=diamond] table [x=snr, y=rl, col sep=comma] {figs/awgn_err_vs_snr.csv};

	  \addlegendentry{Supervised training}
	  \addlegendentry{Alternating training}

		\end{axis}

		\end{tikzpicture}
	    \caption{AWGN channel}
	    \label{fig:awgn_err_vs_snr}
	\end{subfigure} \qquad
    \begin{subfigure}{0.45\linewidth}
	\begin{tikzpicture}
	  \begin{axis}[
	    ymode=log,
	    grid=both,
	    grid style={line width=.1pt, draw=gray!10},
	    major grid style={line width=.2pt,draw=gray!50},
	    minor tick num=5,
	    xlabel={SNR (dB)},
	    ylabel={Error Rate},
	    legend style={at={(0.1, 0.1)},anchor=south west}
	  ]
	    \addplot[orange, mark=triangle] table [x=snr, y=sl, col sep=comma] {figs/rbf_err_vs_snr.csv};
	    \addplot[blue, mark=diamond] table [x=snr, y=rl, col sep=comma] {figs/rbf_err_vs_snr.csv};

	  \addlegendentry{Supervised training}
	  \addlegendentry{Alternating training}

		\end{axis}

		\end{tikzpicture}
	    \caption{\gls{RBF} channel}
	    \label{fig:rbf_err_vs_snr}
	\end{subfigure}

	\caption{Error rate achieved by the alternating and supervised approaches on \gls{AWGN} and \gls{RBF} channels}
	\label{fig:err_vs_snr}
\end{figure*}

The evolutions of the error rates of both the supervised and the alternating approach during the first 500 training iterations are shown in Fig.~\ref{fig:training}, averaged over 200 seeds. Shaded areas around the curves correspond to one standard deviation in each direction.
For the \gls{AWGN} channel, the supervised method leads to faster convergence compared to the alternating method. This is expected since exact gradients are pushed from the receiver to the transmitter through the simulated communication channel. This provides more fine-grained feedback to the transmitter as compared to the \gls{RL}-based approach which provides only feedback of the raw losses (from which a noisy gradient is estimated). However, this can only be done if a differentiable model of the communication medium is available at training, that approximates well the real channel. After approximately 400 training iterations, no significant performance difference is observed.
Surprisingly, for the \gls{RBF} channel, the alternating method enables faster convergence, and with significantly less variance. This can be explained by the fact that, when pushing the gradient through the channel as done with the supervised approach, the gradient is \emph{directly} impacted by the channel response, which is random and therefore leads to high gradient variance. However, with the alternating method, the per-example losses provided to the transmitter are less impacted by the random channel response.
Fig.~\ref{fig:err_vs_snr} shows the achieved error rates for both training methods on the \gls{AWGN} (Fig.~\ref{fig:awgn_err_vs_snr}) and \gls{RBF} channel (Fig.~\ref{fig:rbf_err_vs_snr}). This confirms that the alternating training method without channel model enables similar performance to that of the fully supervised approach.

\section{Conclusion and Ongoing Work}

We have presented a novel method to train fully differentiable communications systems from end-to-end  and showed that it achieves similar performance to the fully supervised approach in \cite{8054694}. However, our method does not require any mathematical model of the channel and can therefore be applied to any type of channel without prior analysis.
Our algorithm currently requires an additional reliable channel during training to feedback losses from the receiver to the transmitter. We are currently working on a training scheme which does not require such a dedicated feedback channel.
Other future investigations include optimizing the introduced scheme using more advanced \gls{RL} techniques to possibly increase the convergence speed.
The proposed method can be easily extended to other tasks than error-free communication of messages, e.g., by including source coding and considering end-to-end reconstruction of raw data (e.g., images) or end-to-end classification of raw data observed on remote devices. 

\appendix[Background on Reinforcement Learning]
\label{sec:rl}

\gls{RL} aims to optimize the behavior of \emph{agents} that interact with an \emph{environment} by taking \emph{actions} in order to  minimize a loss.
An agent in a state $s \in \Sc$, takes an action $a \in \Ac$ according to some policy $\pi$. After taking an action, the agent receives a per-example loss $l$. The expected per-example loss given a state and an action is denoted by $\Lc(s,a)$, i.e., $\Lc(s,a) = \EE\left[l \rvert s, a\right]$. $\Lc$ is assumed to be unknown, and the aim of the agent is to find a policy which minimizes the per-example loss.

During the agent's training, the policy $\pi$ is usually chosen to be stochastic, i.e., $\pi(\cdot|s)$ is a probability distribution over the action space $\Ac$ conditional on a state $s$. Using a stochastic policy enables exploration of the agent's environment, which is fundamental in \gls{RL}. Indeed, training in \gls{RL} is similar to a try-and-fail process: the agent takes an action chosen according to its state, and afterwards improves its policy according to the loss received from the environment. Using a stochastic policy, the agent aims to minimize the loss $J(s, \pi)$, defined as
\begin{equation}
  \label{eq:rl_obj}
  J(s, \pi) = \int_{a \in \Ac} \pi(a | s) \Lc(s, a) \,da.
\end{equation}
\balance
Policy gradient methods are considered in this work, in which the agent optimizes a parametric policy $\pi_{\psiv}$~\cite{NIPS1999_1713}, $\psiv$ being the set of parameters.
The agent optimizes the policy by gradient descent on the loss $J$ with respect to $\psiv$, which requires estimating the gradient of $J$ with respect to $\psiv$:
\begin{align}\label{eq:exchange-ex-grad}
  \nabla_{\psiv} J(s, \pi_{\psiv}) &= \int_{a \in \Ac} \mathcal{L}(s, a) \nabla_{\psiv} \pi_{\psiv}(a | s) \,da\notag\\
                    &= \int_{a \in \Ac} \pi_{\psiv}(a | s) \mathcal{L}(s, a) \nabla_{\psiv} \log{\pi_{\psiv}(a|s)} \,da \notag\\
                    &= \EE_{ \pi_{\psiv} }\left[\mathcal{L}(s, a) \nabla_{\psiv}  \log{\pi_{\psiv}(a | s)} \bigg\rvert s \right] 
\end{align}
where the second equality follows from $\nabla \log{\uv} = \frac{\nabla \uv}{\uv}$.
The exchange of integration and differentiation is valid provided some regularity conditions, discussed, for example, in~\cite{l1995note}.


\bibliographystyle{IEEEtran}
\bibliography{IEEEabrv,bibliography}
\pagebreak
\end{document}